\documentclass[a4paper,11pt]{article}
\usepackage{pos}
\usepackage{graphicx}
\usepackage{amsmath}
\usepackage[export]{adjustbox}
\usepackage{subcaption}
\usepackage{float}

\title{VERITAS and HAWC observations of unidentified source LHAASO J2108+5157}
 \ShortTitle{LHAASO J2108+5157 study with VERITAS }

\author*[a]{Sajan Kumar}
\author[a]{for the VERITAS Collaboration}

\affiliation[a]{Department of Physics, University of Maryland \\
, College Park, MD 20742-4111, USA}

\author[b]{Michael Martin}
\author[b]{Xiaojie Wang}
\affiliation[b]{Michigan Technological University, Department of Physics, \\
  1400 Townsend Drive, Houghton, USA}

\onbehalf{for the HAWC collaboration} 

\emailAdd{skumar86@umd.edu}
\emailAdd{avamarti@mtu.edu}

\abstract{Understanding the complete nature of Galactic sources that accelerate cosmic rays up to $10^{15}$ eV energy (Galactic PeVatrons) is still an unsolved problem in high-energy astrophysics. Although supernova remnants have long been considered as the best candidates for Galactic PeVatrons, a clear association of SNRs with PeVatrons needs further exploration. Recently, the LHAASO collaboration published its first catalog of 90 very high energy (VHE) gamma-ray sources, and a few of them have no obvious counterparts at other wavelengths. Here, we will present morphology and spectral analysis of one such unassociated source LHAASO J2108+5157 using VERITAS and HAWC data.
}

\ConferenceLogo{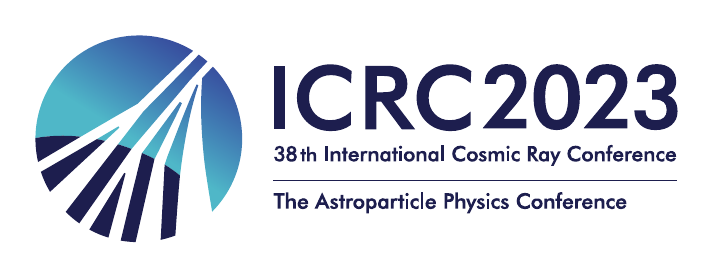}

\FullConference{%
38th International Cosmic Ray Conference (ICRC2023)\\
  26 July - 3 August, 2023\\
  Nagoya, Japan}

\begin{document}
\maketitle

\section{Introduction}
Gamma rays, and in particular at ultra-high energy (UHE; $> 10^{14}$ eV energies), are considered to be a useful probe to understand the nature of Galactic PeVatrons \cite{Bose2022}. 
The exploration of UHE gamma-ray sky started with the detection of Crab Nebula above 100 TeV by Tibet-AS$\gamma$ \cite{Tibet2019uhe}. A year later, High-Altitude Water Cherenkov Observatory (HAWC) published their first UHE catalog, reporting 3 more sources above 100 TeV \cite{HAWCuhe}. These observations made the ground for further exploration of UHE gamma-ray sources. The real thrust to the UHE regime was provided by Large High Altitude Air Shower Observatory (LHAASO) collaboration, when their first catalog of 12 sources was published \cite{Lhaaso2021catalogue}. LHAASO has recently expanded its catalog to around 90 sources, of which 43 are detected above 100 TeV \cite{Lhaaso2023catalogue}. Although the recent advancement in ground-based gamma-ray astronomy, especially with HAWC and LHAASO,  made it possible to detect gamma rays above 100 TeV energies, the real question regarding the nature of object (i.e. Supernova remnants (SNRs), Pulsar and Pulsar Wind Nebulae (PWNs), Pulsar halos, Young Massive Star Clusters (YMSC)) producing these gamma rays is still far from a conclusive answer. Therefore, to understand the nature of these objects, it is critical to follow up the UHE gamma-ray sources at lower energy $\gamma$ rays with imaging atmospheric Cherenkov telescopes (IACTs) such as VERITAS, HESS, and MAGIC. Since the IACTs can resolve the UHE sources with $\le 0.1^{\circ}$ angular resolution, they offer a complimentary view and can shed more light on the gamma-ray source identification.

LHAASO J2108+5157 is one such source detected by LHAASO collaboration in the energy range of 1-25 TeV using Water Cherenkov Detector Array (WCDA) and above 25 TeV using Kilometer Squared Array (KM2A) detector \cite{Lhaaso2021catalogue, LhaasoJ2108-2021, Lhaaso2023catalogue}. However, the non-detection of a counterpart in any other wavelength makes it an exciting candidate to explore further. The power-law index is reported to be changed from $1.44 \pm 0.08$ in 1-25 TeV range to $2.97 \pm 0.07$ above 25 TeV energy (Table 1 in \cite {Lhaaso2023catalogue}). Initially LHAASO J2108+5157 was detected as a point source by only using KM2A detector \cite{LhaasoJ2108-2021}. However, with larger dataset from KM2A and inclusion of WCDA detector, the source is reported to be marginally extended at a value of $0.19 \pm 0.02$ deg and $0.14 \pm 0.03$ deg in KM2A and WCDA data respectively \cite{Lhaaso2023catalogue}. The Large-Sized Telescope prototype (LST-1) has also observed this source for 49 hours, however, no detection is reported. The differential spectrum above 300 GeV using $95\%$ confidence upper limits is described with a hard power-law with a spectral index of $1.62 \pm 0.23$ by LST-1 \cite{LST2023}. A dedicated 12.2 years of Fermi-LAT analysis of the region around LHAASO J2108+5157 detected a hard spectrum (HS) source (at $4\sigma$ level, photon index $1.9 \pm 0.2$) \cite{LST2023} in addition to previously detected soft spectrum source 4FGL J2108+5155 (No emission above 2 GeV) \cite{LhaasoJ2108-2021}. This HS source is separated from LHAASO J2108+5157 at a distance of $\sim 0.27$ deg. Since this distance is larger than the extension upper limit reported in \cite{LhaasoJ2108-2021}, this HS is unlikely to be associated with LHAASO J2108+5157.   
Similarly, 4FGL J2108+5155 shows a steep spectrum above a few GeV, which makes it incompatible with LST-1 and LHAASO spectral measurements. Moreover, no significant X-ray emission has been detected in the Swift-XRT survey of the LHAASO J2108+5157 region and also no energetic pulsar is found in the nearby region; the closest X-ray source is the binary RX J2107.3$+$5202.

As no powerful pulsars or supernova remnants has been detected so far in the vicinity of LHAASO J2108+5157, it is difficult to conclude the origin of gamma-ray emission. A multiwavelgnth modelling of gamma-ray emission from LHAASO J2108+5157 can be described by either hadronic and leptonic models. Recently, a hadronic scenario has been proposed where cosmic rays escaped from an old supernova remnant and interacting with near by molecular clouds produced the gamma rays \cite{Sarkar2023, J2108-MC-2023}. Similarly, pulsar-like spectral signature of 4FGL J2108+5155 makes leptonic scenario also plausible to explain the gamma-ray emission from LHAASO J2108+5157 \cite{LhaasoJ2108-2021, LST2023}.

\section{VERITAS data and analysis}
The VERITAS observatory is situated in southern Arizona
($31.68 \mathrm{\ N}$, $110.95\mathrm{\ W}$) at an elevation of $1268 \mathrm {\ m}$ above sea level. It consists of an array of four air shower Cerenkov telescopes, each with an effective mirror diameter of $12 \ \mathrm{m}$. The camera of each telescope consists of 499 Photo Multiplier Tubes (PMT) resulting in a field of view of $3.5^{\circ}$ in the sky. VERITAS is designed to detect gamma rays in the energy range from $85 \ \mathrm{GeV}$ to more than $30 \ \mathrm{TeV}$. With its current sensitivity, a source with a flux level of 1\% of steady flux from the Crab Nebula can be detected in 25h. The angular resolution of the array at 1 TeV is $\sim 0.1^{\circ}$. A detailed description of the VERITAS performance is given in \cite{Nahee2015}.

VERITAS observed LHAASO J2108+5157 for 40 hours in 2021, which was reduced to 35 hours of good quality data after applying quality cuts. The data were analyzed using the EventDisplay package \cite{ED2017}. A minimum of two telescopes are required for the event reconstruction. As most of the events which trigger our telescopes belong to background events, we removed these background events using the machine learning classification method \textit{boosted decision tree} discussed in \cite{Maria2017}. The reconstruction and event selection results in an energy threshold of $300 \ \mathrm{GeV}$ for our analysis. Although the classification algorithm can remove more than $99\%$ of the background events from our sample of events, there is still an irreducible background, which was estimated using the ring background method \cite{Berge2007}.

\section{HAWC data and analysis}
High Altitude Water Cherenkov (HAWC) gamma-ray observatory is a ground-based water Cherenkov instrument located at Sierra Negra, Mexico, at an altitude of 4,100 meters. It consists of 300 tanks in the main array resulting in an area of $22000 \ \mathrm{m^{2}}$. Each tank is equipped with three 8-inch Hamamatsu photomultiplier tubes (PMTs) and one 10-inch high-quantum efficiency Hamamatsu PMT. HAWC is sensitive to Extensive Air Shower (EAS) events from 300~GeV to above 100~TeV. It has a duty cycle of greater than 95~\%, which could be a great connection and complement to the IACT experiments.

The HAWC data for $\sim $2400~days has been analyzed for the LHAASO J2108+5157 region using the \textit{Pass5} reconstruction. 
The data are binned into 2D bins based on the triggering fraction of the array and the reconstructed energy of the events. The details about the analysis can be found here in \cite{hawc2019crab}.

\section{Results}
\begin{figure}[h]
 
\begin{subfigure}{0.5\textwidth}
\includegraphics[scale=0.6]{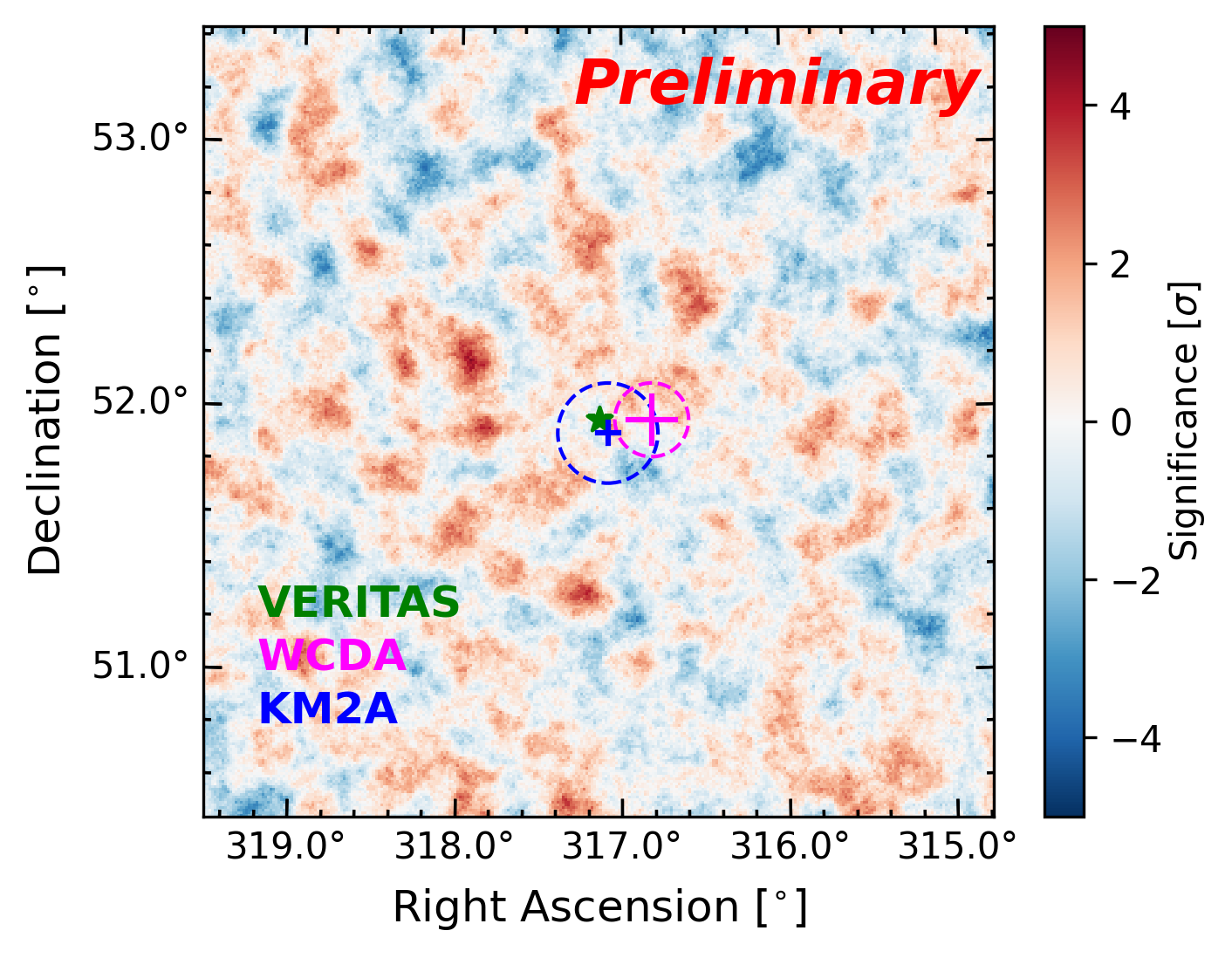} 
\label{fig:subim1}
\vspace{1.9mm}
\end{subfigure}
\begin{subfigure}{0.4\textwidth}
\includegraphics[scale=0.08]{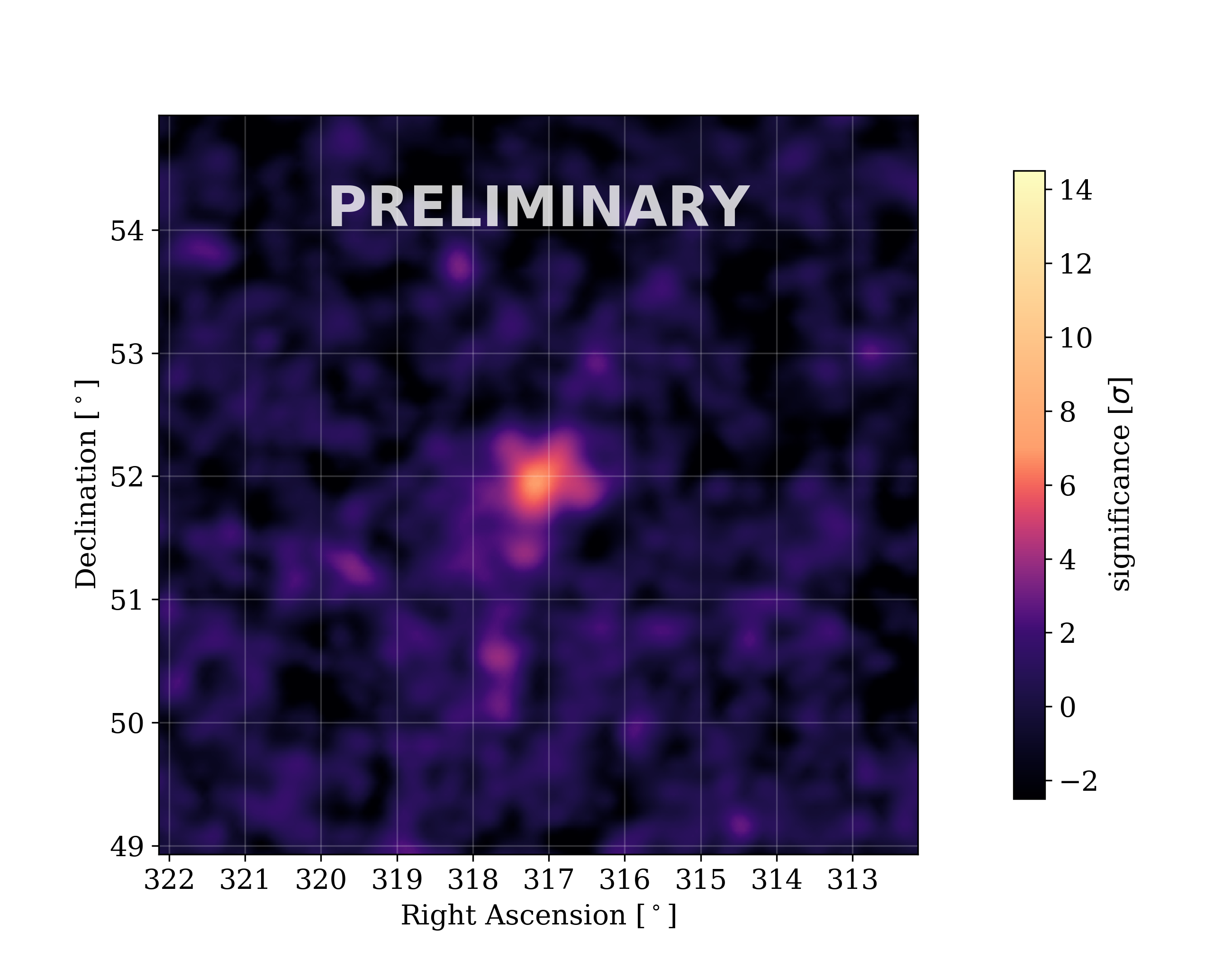}
\label{fig:subim2}
\end{subfigure}
 
\caption{\textit{Left}: VERITAS significance map created using point source analysis (integration radius = $0.09^{\circ}$). Green star indicates the VERITAS location around which significance of source is estimated ($0.6 \sigma$). Magenta and blue crosses represents the position of LHAASO-WCDA and LHAASO-KM2A with $95\%$ error respectively, with the dashed circles representing the corresponding one-sigma extension of a 2D Gaussian model \cite{Lhaaso2023catalogue}. \textit{Right}: Significance map of J2108 region using $\sim 2400$ days of data taken by HAWC.}
\label{fig:image2}
\end{figure}

\begin{figure}[h]
\begin{center} 
  \includegraphics[scale=0.6]{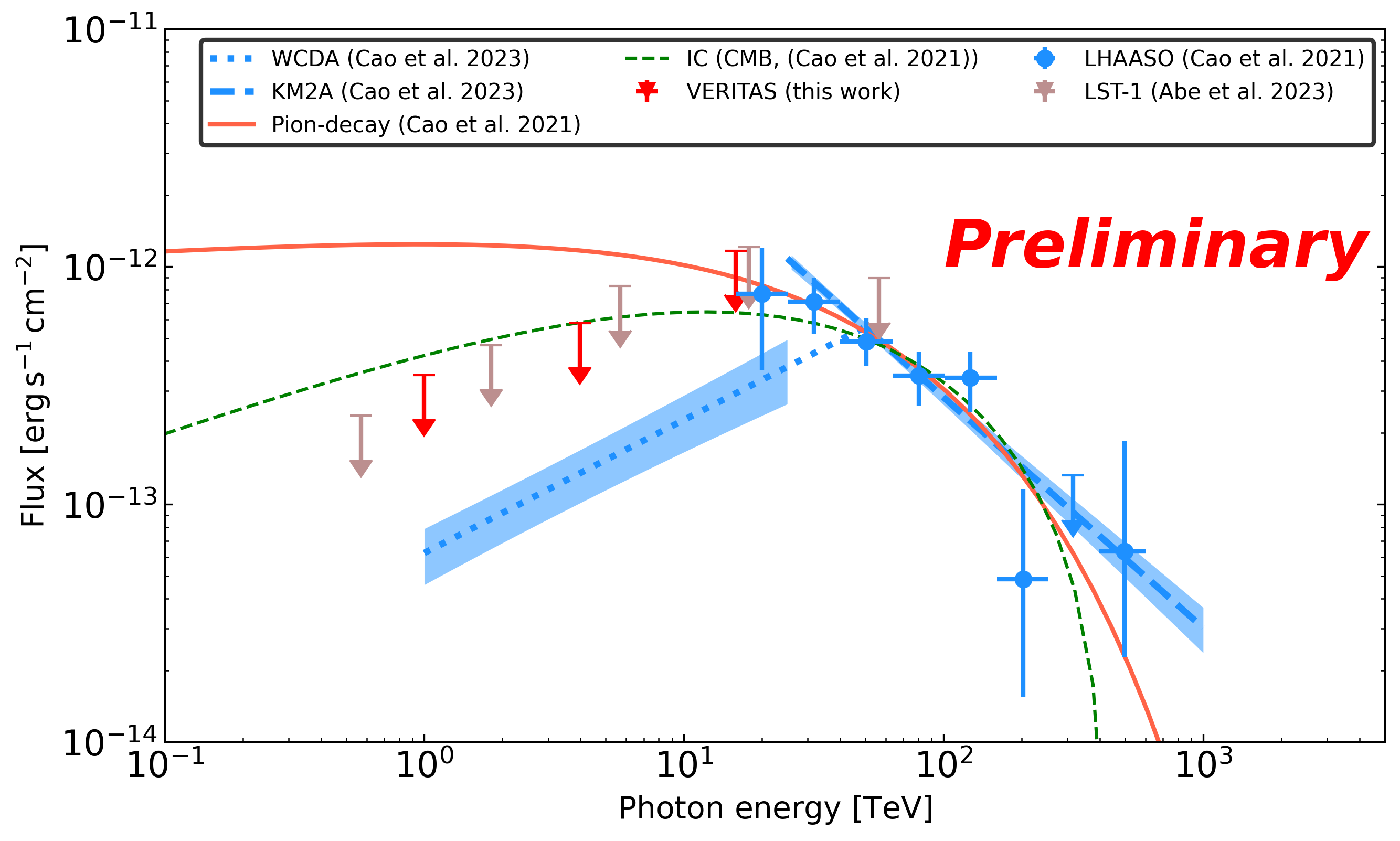}
  \caption{VERITAS spectral upper limits are shown along with spectral measurements from LST-1 \cite{LST2023} and LHAASO \cite{Lhaaso2023catalogue, LhaasoJ2108-2021}. It should be noted that the VERITAS ULs are not directly comparable with LHAASO-WCDA results since the VERITAS limits are extracted from an integration radius of $0.09^{\circ}$, whereas WCDA spectrum is produced using 2D Gaussian model with $r_{39} = 0.14^{\circ}$ ($r_{99} = 0.42^{\circ}$). Moreover, there is also a position offset of $0.2^{\circ}$ between VERITAS and WCDA (see Figure \ref{fig:image2}).}
  \label{fig:spectrum2108}
\end{center}
\end{figure}

Figure \ref{fig:image2} (left) shows the significance map of LHAASO J2108+5157 region using VERITAS data. The map is created using an integration radius of $0.09^{\circ}$ (consistent with VERITAS PSF) \cite{LhaasoJ2108-2021, LST2023}. The significance at the centroid location reported by LHAASO \cite{LhaasoJ2108-2021} ((RA = $317.15^{\circ}$, Dec = $51.95^{\circ}$)) is estimated at $0.6\sigma$ level. Therefore, we are reporting a non-detection of gamma-ray emission from LHAASO J2108+5157. Spectral analysis is performed on the same circular region of radius $0.09^{\circ}$ around the position of LHAASO J2108+5157. The resulting upper limits at $95\%$ confidence level are shown in Figure \ref{fig:spectrum2108}.

Figure \ref{fig:image2} (right) shows the significance map of the LHAASO J2108+5157 region based on approximately 2400 days of HAWC observations. The map reveals a maximum significance of 7$\sigma$ at coordinates RA=317.15$^\circ$ and Dec=51.93$^\circ$. The significance plot is obtained assuming a point source with a power-law spectrum having an index of -2.6. To accurately describe the gamma-ray emission detected by HAWC, various morphology and spectrum models are being tested. Detailed analysis results, which are currently being finalized, will be released in the near future.

\section{Conclusion}
No significant emission close to the position of LHAASO J2108+5158 has been detected with VERITAS data. Therefore, differential flux upper limits at $95\%$  confidence level are measured at 1.0, 3.98, and 15.38 TeV energy. These ULs are consistent with LST-1 results \cite{LST2023} as shown in Figure \ref{fig:spectrum2108}.

The absence of known pulsars or a supernova remnant makes it challenging to understand the elusive nature of LHAASO J2108+5157. The upper limits from VERITAS and LST-1 \cite{LST2023} exclude the hadronic model described in \cite{LhaasoJ2108-2021} (spectral index is harder than assumed in \cite{LhaasoJ2108-2021}) and hints towards a leptonic origin of emission from few TeV to hundreds of TeV energy. However, more recently, a new molecular cloud with a nucleon density of $133 \ \mathrm{cm^{-3}}$ and at a distance of $1.6\pm0.1 \ \mathrm{kpc}$  has been found in the vicinity of LHAASO J2108+5157  \cite{J2108-MC-2023}. The morphology of this new cloud highly correlates with the LHAASO J2108+5157 gamma-ray emission up to $2 \ \mathrm{GeV}$ from \textit{Fermi}-LAT and emission detected by LHAASO. This makes it more likely that the gamma rays are produced through the hadronic channel with molecular cloud as the main target for the cosmic ray particles accerlerated by an unidentified PeVatrons \cite{Sarkar2023, J2108-MC-2023}. As it is very difficult to distinguish between leptonic and hadronic emisson at the highest energies, future observations by CTA (with an order of magnitude better sensitivity) and analysis in the x-ray band (constraining magnetic field) will be helpful to identify true nature of this PeVatron.

\section*{Acknowledgements}
This research is supported by grants from the U.S. Department of Energy Office of Science, the U.S. National Science Foundation and the Smithsonian Institution, and by NSERC in Canada. This research used resources provided by the Open Science Grid, which is supported by the National Science Foundation and the U.S. Department of Energy’s Office of Science, and resources of the National Energy Research Scientific Computing Center (NERSC), a U.S. Department of Energy Office of Science User Facility operated under Contract No. DE-AC02-05CH11231. We acknowledge the excellent work of the technical support staff at the Fred Lawrence Whipple Observatory and at the collaborating institutions in the construction and operation of the instrument.

We acknowledge the support from: the US National Science Foundation (NSF); the US Department of Energy Office of High-Energy Physics; the Laboratory Directed Research and Development (LDRD) program of Los Alamos National Laboratory; Consejo Nacional de Ciencia y Tecnolog\'{i}a (CONACyT), M\'{e}xico, grants 271051, 232656, 260378, 179588, 254964, 258865, 243290, 132197, A1-S-46288, A1-S-22784, CF-2023-I-645, c\'{a}tedras 873, 1563, 341, 323, Red HAWC, M\'{e}xico; DGAPA-UNAM grants IG101323, IN111716-3, IN111419, IA102019, IN106521, IN110621, IN110521 , IN102223; VIEP-BUAP; PIFI 2012, 2013, PROFOCIE 2014, 2015; the University of Wisconsin Alumni Research Foundation; the Institute of Geophysics, Planetary Physics, and Signatures at Los Alamos National Laboratory; Polish Science Centre grant, DEC-2017/27/B/ST9/02272; Coordinaci\'{o}n de la Investigaci\'{o}n Cient\'{i}fica de la Universidad Michoacana; Royal Society - Newton Advanced Fellowship 180385; Generalitat Valenciana, grant CIDEGENT/2018/034; The Program Management Unit for Human Resources \& Institutional Development, Research and Innovation, NXPO (grant number B16F630069); Coordinaci\'{o}n General Acad\'{e}mica e Innovaci\'{o}n (CGAI-UdeG), PRODEP-SEP UDG-CA-499; Institute of Cosmic Ray Research (ICRR), University of Tokyo. H.F. acknowledges support by NASA under award number 80GSFC21M0002. We also acknowledge the significant contributions over many years of Stefan Westerhoff, Gaurang Yodh and Arnulfo Zepeda Dominguez, all deceased members of the HAWC collaboration. Thanks to Scott Delay, Luciano D\'{i}az and Eduardo Murrieta for technical support.

\bibliographystyle{unsrt}
\bibliography{references.bib}


\clearpage


\begin{center} 
\section*{\centering{All Authors and Affiliations}}

\normalsize{VERITAS COLLABORATION  \\}

\scriptsize
\noindent
A.~Acharyya$^{1}$,
C.~B.~Adams$^{2}$,
A.~Archer$^{3}$,
P.~Bangale$^{4}$,
J.~T.~Bartkoske$^{5}$,
P.~Batista$^{6}$,
W.~Benbow$^{7}$,
J.~L.~Christiansen$^{8}$,
A.~J.~Chromey$^{7}$,
A.~Duerr$^{5}$
% ...

\vspace{1cm}


\noindent
$^{1}${Department of Physics and Astronomy, University of Alabama, Tuscaloosa, AL 35487, USA}

$^{2}${Physics Department, Columbia University, New York, NY 10027, USA}

$^{3}${Department of Physics and Astronomy, DePauw University, Greencastle, IN 46135-0037, USA}

$^{4}${Department of Physics and Astronomy and the Bartol Research Institute, University of Delaware, Newark, DE 19716, USA}

$^{5}${Department of Physics and Astronomy, University of Utah, Salt Lake City, UT 84112, USA}

$^{6}${DESY, Platanenallee 6, 15738 Zeuthen, Germany}

$^{7}${Center for Astrophysics $|$ Harvard \& Smithsonian, Cambridge, MA 02138, USA}

$^{8}${Physics Department, California Polytechnic State University, San Luis Obispo, CA 94307, USA}

% ...

\end{center}


\end{document}